\documentclass[twocolumn,secnumarabic,amssymb, nobibnotes, aps, prc]{revtex4-1}
\usepackage{graphicx,amsmath,amsthm}

\begin{document}

\title{High precision probe of the fully sequential decay width of the Hoyle state in $^{12}$C}

\author{D. Dell'Aquila\textsuperscript{1,2,3}}
\thanks{\underline{dellaquila@na.infn.it}}
\author{I. Lombardo\textsuperscript{1,4}}
\thanks{\underline{ivlombardo@na.infn.it}}
\author{G. Verde\textsuperscript{3,4}} 
\author{M. Vigilante\textsuperscript{1,2}}
\author{L. Acosta\textsuperscript{5}} 
\author{C. Agodi\textsuperscript{6}} 
\author{F. Cappuzzello\textsuperscript{7,6}} 
\author{D. Carbone\textsuperscript{6}} 
\author{M. Cavallaro\textsuperscript{6}} 
\author{S. Cherubini\textsuperscript{6,7}}
\author{A. Cvetinovic\textsuperscript{6}} 
\author{G. D'Agata\textsuperscript{7,6}} 
\author{L. Francalanza\textsuperscript{2}} 
\author{G.L. Guardo\textsuperscript{6}} 
\author{M. Gulino\textsuperscript{8,6}} 
\author{I. Indelicato\textsuperscript{6}} 
\author{M. La Cognata\textsuperscript{6}} 
\author{L. Lamia\textsuperscript{7}} 
\author{A. Ordine\textsuperscript{2}} 
\author{R.G. Pizzone\textsuperscript{6}} 
\author{S.M.R. Puglia\textsuperscript{6}} 
\author{G.G. Rapisarda\textsuperscript{6}} 
\author{S. Romano\textsuperscript{6}} 
\author{G. Santagati\textsuperscript{6}} 
\author{R. Spart\`a\textsuperscript{6}}
\author{G. Spadaccini\textsuperscript{1,2}} 
\author{C. Spitaleri\textsuperscript{7,6}} 
\author{A. Tumino\textsuperscript{8,6}}

\affiliation{$^1$ Dip. di Fisica "E. Pancini", Universit\`{a} di Napoli Federico II, I-80126 Napoli, Italy}
\affiliation{$^2$ INFN-Sezione di Napoli, I-80126 Napoli, Italy} 
\affiliation{$^3$ Institut de Physique Nucl\'{e}aire, CNRS-IN2P3, Univ. Paris-Sud, Universit\'{e} Paris-Saclay, 91406 Orsay Cedex, France} 
\affiliation{$^4$ INFN - Sezione di Catania, Via S. Sofia, I-95125 Catania, Italy}
\affiliation{$^5$ Universidad Nacional Aut\'{o}noma de M\'{e}xico, A.P. 20-364, Cd.Mx, D.F. 01000 M\'{e}xico}
\affiliation{$^6$ INFN - Laboratori Nazionali del Sud, Via S. Sofia, I-95125 Catania, Italy}
\affiliation{$^7$ Dip. di Fisica e Astronomia, Universit\`{a} di Catania, Via S. Sofia, I-95125 Catania, Italy}
\affiliation{$^8$ Facolt\`{a} di Ingegneria ed Architettura, Universit\`{a} Kore, I-94100 Enna, Italy}

\date{\today}%

\begin{abstract}
The decay path of the Hoyle state in $^{12}$C ($E_x=7.654\textrm{MeV}$) has been studied with the $^{14}\textrm{N}(\textrm{d},\alpha_2)^{12}\textrm{C}(7.654)$ reaction induced at $10.5\textrm{MeV}$. High resolution invariant mass spectroscopy techniques have allowed to unambiguously disentangle direct and sequential decays of the state passing through the ground state of $^{8}$Be. Thanks to the almost total absence of background and the attained resolution, a fully sequential decay contribution to the width of the state has been observed. The direct decay width is negligible, with an upper limit of $0.043\%$ ($95\%$ C.L.). The precision of this result is about a factor $5$ higher than previous studies. This has significant implications on nuclear structure, as it provides constraints to $3$-$\alpha$ cluster model calculations, where higher precision limits are needed.
\end{abstract}
\maketitle

Exploring the structure of $^{12}$C is extremely fascinating, since it is strongly linked to the existence of $\alpha$ clusters in atomic nuclei and to the interplay between nuclear structure and astrophysics. Furthermore, $^{12}$C is one of the major constituents of living beings and ourselves. Our present knowledge traces the origin of $^{12}$C to the so called $3\alpha$ process in stellar nucleosynthesis environments. The $3\alpha$ process, which occours in the He-burning stage of stellar nucleosynthesis, proceeds via the initial fusion of two $\alpha$ particles followed by the fusion with a third one \cite{opik,salpeter} and the subsequent radiative de-excitation of the so formed excited carbon-12 nucleus, $^{12}$C$^*$. The short lifetime of the $^{8}$Be unbound nucleus (of the order of $10^{-16}s$), formed in the intermediate stage, acts as a bottle-neck for the whole process. Consequently, the observed abundance of carbon in the universe cannot be explained by considering a non-resonant two-step process. This fact led Fred Hoyle, in 1953, to the formulation of his hypothesis \cite{hoyle_first,hoyle}: the second step of the $3\alpha$ process, $\alpha + ^{8}$Be$\rightarrow ^{12}$C$ + \gamma$, has to proceed through a resonant $J^{\pi}=0^+$ state in $^{12}$C, close to the $\alpha+^{8}$Be emission threshold. The existence of such a state was then soon confirmed \cite{cook} at an excitation energy of $7.654\textrm{MeV}$. This state was then named as the \emph{Hoyle state} of $^{12}$C \cite{freer_hoyle_review}.

The decay properties of this state strongly affect the creation of carbon and heavier elements in helium burning \cite{ogata}, as well as the evolution itself of stars \cite{Herwig,Tur:2009zb}. At typical stellar temperatures of $T\approx10^8-10^9\textrm{K}$, this reaction proceeds exclusively via \textit{sequential} process consisting of the $\alpha+\alpha$ s-wave fusion to the ground state of $^{8}$Be, followed by the s-wave radiative capture of a third $\alpha$ to the Hoyle state. However, in astrophysical scenarios that burn helium at lower temperatures, like for instance helium-accreting white dwarfs or neutron stars with small accretion rate, another decay mode of the Hoyle state completely dominates the reaction rate: the non-resonant, or \emph{direct}, $\alpha$ decay \cite{nguyen,nomoto,langanke}, where the two $\alpha$s bypass the formation of $^{8}$Be via the $92\textrm{keV}$ resonance. Recent theoretical calculations show that, at temperatures below $0.07\textrm{GK}$, the reaction rate of the direct process is largely enhanced with respect to the one calculated by assuming only the sequential scenario \cite{nacre}; as an example, for temperatures around $0.02\textrm{GK}$ such enhancement is predicted to be $7$-$20$ orders of magnitude \cite{nguyen,nguyenprl,garrido,yabana,ogata}.

In nuclear structure, the Hoyle state is crucial to understand clustering in nuclei \cite{vonoertzen,uegaki,kamimura}. Theoretical calculations show different hypothesis regarding its spatial configuration. Recent ab-initio calculations describe it as a gas-like diluted state \cite{uegaki,kamimura}, where the constituent $\alpha$ clusters are only weakly interacting. The possible appearance of Bose-Einstein condensates of $\alpha$ particles have been also proposed \cite{tohsaki,funaki,funaki1}, as well as molecular-like structures with three $\alpha$'s forming a linear chain, an obtuse triangle or a bent-arm configuration \cite{chernykh,kamimura,uegaki,morinaga,epelbaum1}. Between several observables, some of these models are able to predict the sequential-to-direct decay branching ratio (B.R.) of the Hoyle state \cite{ishikawa}. The accurate knowledge of the experimental value of such branching ratio has therefore the capital importance to serve as a benchmark of theoretical models attempting to describe $\alpha$ clustering in $^{12}$C.   

Recently, a quite large number of experiments has been carried out to probe the structure and decay properties of the Hoyle state in $^{12}$C. The most commonly adopted strategy is to explore how the Hoyle state decays via $3\alpha$ emission, i.e. what is \emph{direct} decay rate relative to the \emph{sequential} one. An upper limit to the direct decay branch was firstly given by Freer et al. in 1994 \cite{freer_hoyle1994}. In their work they suggested that the B.R. of the Hoyle state decay bypassing the $^{8}$Be ground state was lower than $4\%$, i.e. $(\Gamma_\alpha-\Gamma_{\alpha_0})/\Gamma_\alpha<0.04$. Here $\Gamma_\alpha$ indicates the global $\alpha$ decay width and $\Gamma_{\alpha_0}$ is the partial width of the $\alpha$ emission leading to the ground state of $^{8}$Be. More recently, Raduta et al.~\cite{raduta} reported a result in strong contradiction with the previous one, finding a rather high value ($17\% \pm 5\%$) of direct B.R. Such contrasting results stimulated a series of new experiments aimed at determining the actual value of the direct decay B.R. of the Hoyle state. A new upper limit of $0.5\%$ ($95\%$ C.L.) was obtained by Kirsebom et al. by using the kinematic fitting method \cite{kirsebom}. Two more recent experiments by Rana et al.~\cite{rana} and Morelli et al.~\cite{morelli_hoyle} suggested non-zero values of direct decay B.R., respectively of $(\Gamma_\alpha-\Gamma_{\alpha_0})/\Gamma_\alpha=0.91\%\pm0.14\%$ and $1.1\%\pm0.4\%$. Finally, thanks to a high statistics experiment, Itoh et al.~\cite{itoh} determined an improved upper limit of the direct B.R. of $0.2\%$ ($95\%$ C.L.). It is important to underline that, as discussed in \cite{freer_hoyle1994,itoh}, the use of strip detectors introduces the presence of a non-vanishing background, that reduces the sensitivity to the direct decay B.R. signal. Taking into account the importance to fully understand $\alpha$ clustering effects in the nuclear structure of $^{12}$C, it is mandatory to improve our knowledge of the direct-to-sequential decay B.R. of the Hoyle state, since theoretical estimations of this quantity are given at the $0.1\%$ level, i.e. well below the most recent upper limit reported in the literature \cite{freer_models,itoh}.

In this letter we report on the result of a new high precision experiment specifically designed to isolate, if any, $3\alpha$ direct decays of the Hoyle state in $^{12}$C. For the first time we succeeded to have almost \emph{zero background}, which is a requirement in order to unambiguously disentangle sequential and direct decays. To populate $^{12}$C nuclei in the Hoyle state we used the $^{14}\textrm{N}(\textrm{d},\alpha)^{12}\textrm{C}$ nuclear reaction. A $10.5\textrm{MeV}$ deuteron beam was provided by the $15MV$ tandem accelerator of the INFN-LNS (Catania, Italy). As a detection apparatus we used the combination of a $\Delta$E-E telescope and a high granularity hodoscope detector. The adopted experimental method is the invariant mass analysis of $3\alpha$ disintegrations of the Hoyle state. We completely reconstruct the kinematics of the reaction by simultaneously detecting the four $\alpha$ particles emitted in the final state, namely the $\alpha$ ejectile, used to tag the excitation of the $^{12}$C residue at its Hoyle state ($E^*=7.654\textrm{MeV}$), and the three $\alpha$ particles fed by the Hoyle state decay. 

The hodoscope detector was specifically designed to ensure the detection of the three $\alpha$ particles coming from the Hoyle state decay with the highest possible efficiency and to avoid the artificial introduction of background. It is constituted by $8\times8$ independent silicon pads ($1cm^2$, 300$\mu$m thick), and it is placed in such a way that its center is aligned with the axis of the $^{12}$C($7.654$) three $\alpha$ emission cone, when the corresponding $\alpha$ tagging ejectile is detected by the $\Delta$E-E telescope.

\begin{center}
	\begin{figure}[t]
		\includegraphics[scale=0.4]{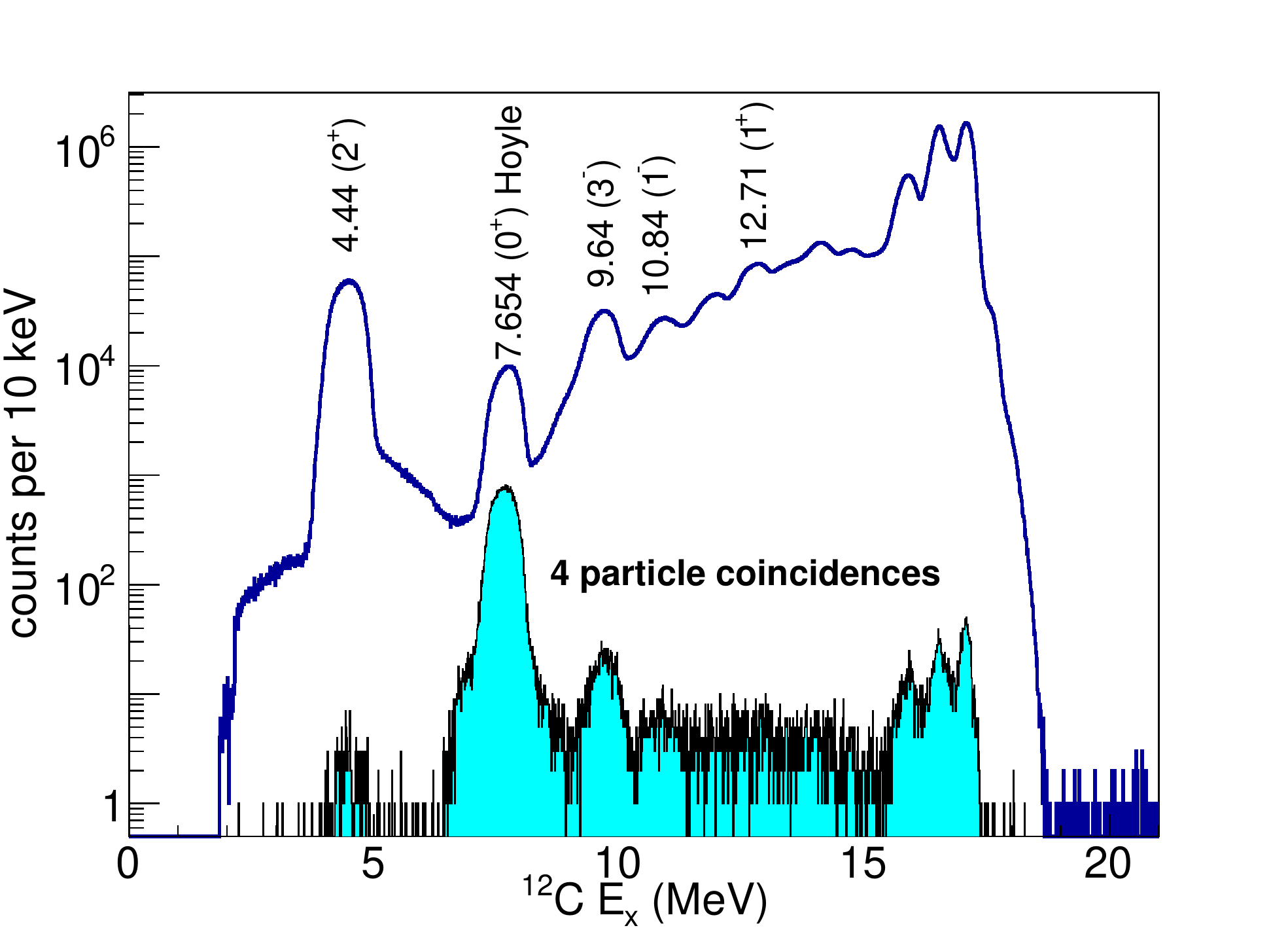}
		\caption{$^{12}$C excitation energy spectrum reconstructed from the measured momentum of particles detected by the $\Delta$E-E anti-coincidence telescope (blue line). Labels are used to indicate the energy and spins of well-known states of $^{12}$C. (filled histogram) Same spectrum obtained by selecting $4$-particles coincidences.}
		\label{fig:AntiTel}
	\end{figure}	
\end{center}  

The $^{12}$C excitation energy spectrum, reconstructed from the measurement of kinetic energy and emission direction of the particles detected in the $\Delta$E-E telescope, is shown on Fig.~\ref{fig:AntiTel} by the blue line. Only particles stopping in the first detection stage are selected, allowing to strongly reduce contaminations from (d,d) and (d,p) reactions on the target constituents. Details on this technique can be found e.g. in Ref.~\cite{koenig}. The excitation energy spectrum reduces to the filled one if we select events with $4$-particles in coincidence, i.e. by selecting events where $3$ particles are detected in coincidence by the hodoscope. This spectrum exhibits a pronounced peak at $E_x = 7.654\textrm{MeV}$, corresponding to the energy position of the Hoyle state, while background as well as other peaks are strongly suppressed, demonstrating the good sensitivity of our detection system to the $\alpha$ decays of the Hoyle state and a very low background level. For the subsequent analysis, events are selected by gating on the Hoyle peak and on the corresponding four-particles total energy spectrum, which unambiguously identifies the reaction channel of interest.

\begin{center}                                                                                                                
	\begin{figure}[t]
		\includegraphics[scale=0.4]{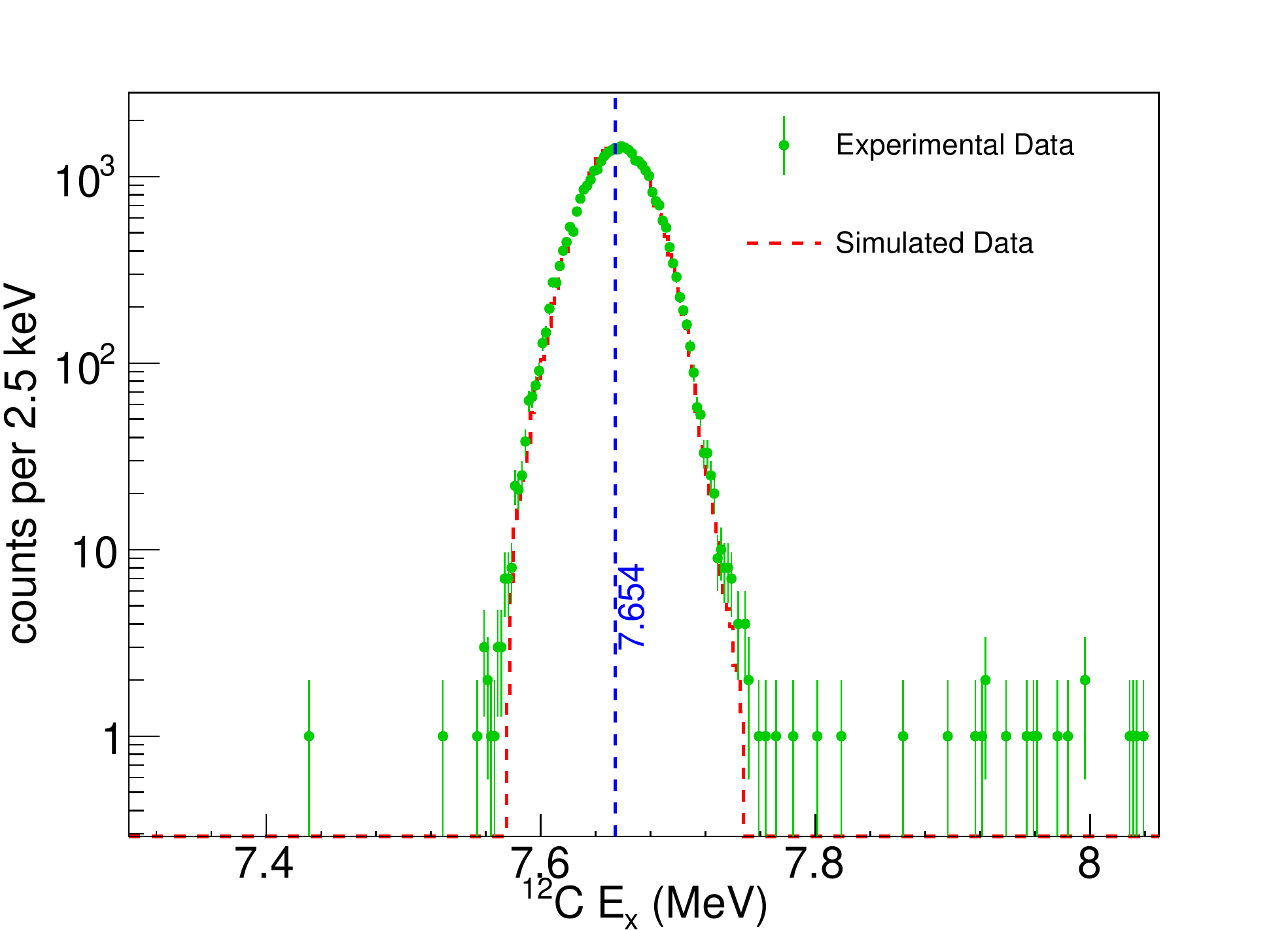}
		\caption{Three particles invariant mass spectrum ($^{12}$C $E_{x}$) gated on the Hoyle peak of Fig.~\ref{fig:AntiTel}. Experimental points are compared with the result of a  Monte Carlo simulation of the $^{14}\textrm{N}(\textrm{d},\alpha_2)^{12}\textrm{C}(7.654)$, which details are explained in the text. Events under the peak centered at $7.654\textrm{MeV}$ are due to the decay of the Hoyle state. The evaluated background level is about $0.036\%$.}
		\label{fig:Ex}
	\end{figure}	
\end{center}  

In Fig.~\ref{fig:Ex} we report (full dots) the $^{12}$C excitation energy spectrum obtained by an invariant mass analysis of ternary coincidences inside the hodoscope, assuming that they are $\alpha$ particles. The red dashed line is the result of a complete Monte Carlo simulation of the effect of the detection system on the reconstruction of the three $\alpha$ particles resulting from the in-flight decay of the Hoyle state. To produce this result we consider four $\alpha$ particles fully reconstructed events from $^{14}\textrm{N}(\textrm{d},\alpha_2)^{12}\textrm{C}(7.654)$ reaction simulated data. In our simulation we have taken into account both the profile of the beam on the target and the angular distribution of the emitted $\alpha$ ejectile, as reported in \cite{curry} at the same incident energy. The geometry of the detectors and their energy resolution are also taken into account in the simulation. The result of the simulation is in excellent agreement with the experimental data, confirming the unambiguous reconstruction of this physical process. 

The invariant mass of the Hoyle state is determined with a resolution of about $47\textrm{keV}$ (FWHM), while the center of the distribution is in agreement with the position of the Hoyle state within an indetermination smaller than $1\textrm{keV}$. Four-$\alpha$ fully detected events are thus selected by means of a further cut on the peak of Fig.~\ref{fig:Ex}. In such a way we obtain a number of about 28000 decay events of Hoyle state, an amount well higher than any other previous investigation. The background level, due to spurious coincidences, is extremely low thanks to the stringent constraints on the data, the sensitivity of the apparatus to the physical process and the unambiguous particle tracks identification achieved by the use of an hodoscope. It can be evaluated by inspecting the right and left sides of the spectrum; it amounts to about $0.036\%$ of the total integral of the peak.

\begin{center}                                                                                                                
	\begin{figure}[t]
		\includegraphics[scale=0.4]{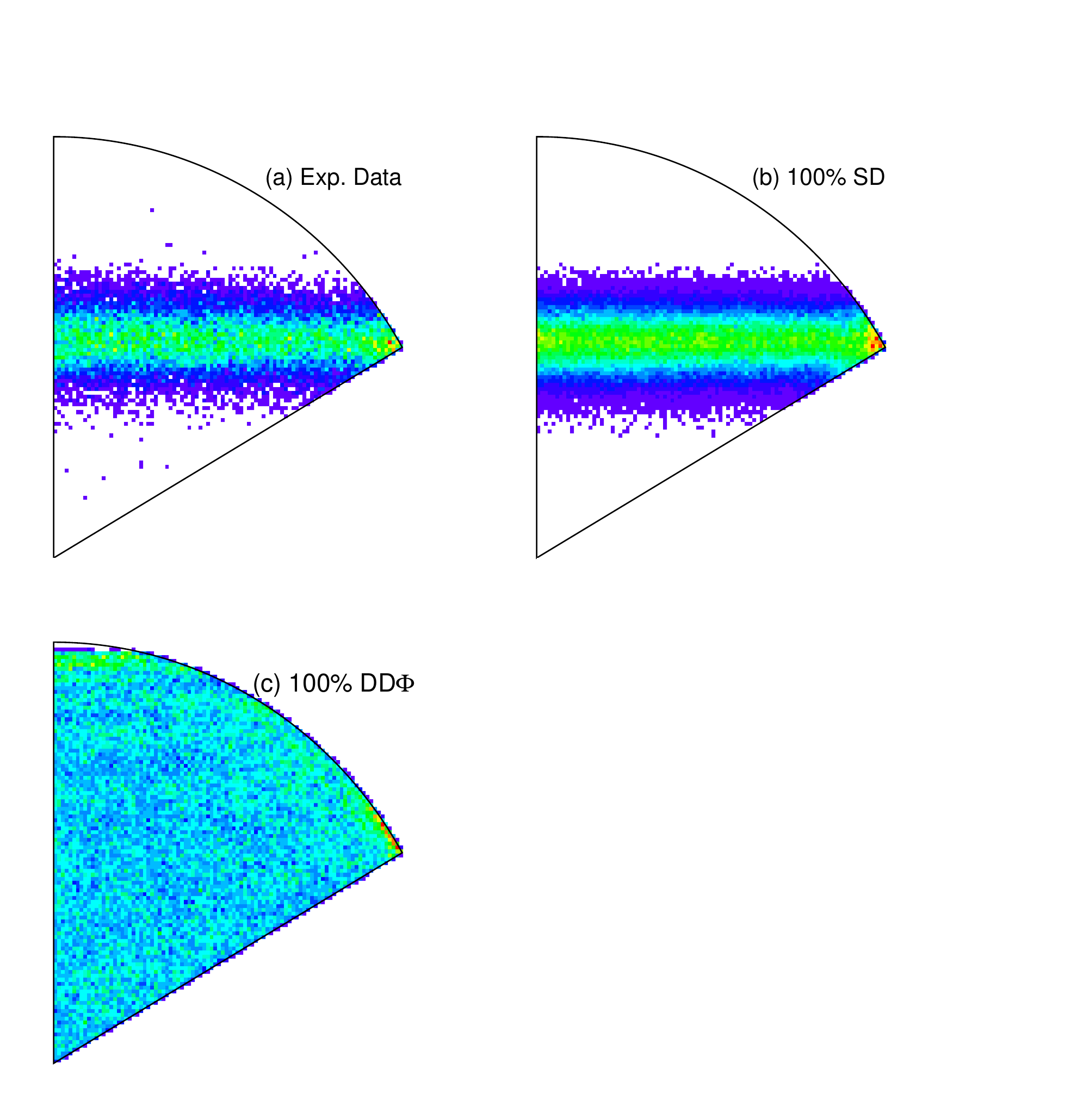}
		\caption{Experimental symmetric Dalitz plot (a) compared with Monte Carlo simulations of (b) $100\%$ SD and (c) $100\%$ DD$\Phi$ decays. Further details on the simulation procedure are explained in the text. Simulated SD events result in an horizontal band and the effect of our experimental apparatus is expected to not introduce any significant contamination outside this band, as instead observed in previous works \cite{itoh,freer_hoyle1994}.}
		\label{fig:dalitz_plot}
	\end{figure}	
\end{center}  

Details about the three-$\alpha$ decay mechanisms of the Hoyle state can be studied by using the symmetric Dalitz plot \cite{dalitz}. This technique is particularly suited to geometrically visualize the decay pattern into three equal mass particles. Cartesian coordinates to construct the Dalitz plot can be obtained as follows:
\begin{equation}
  \begin{split}
    &x=\sqrt{3}(\varepsilon_j-\varepsilon_k) \\
    &y=2\varepsilon_i-\varepsilon_j - \varepsilon_k
  \end{split}
\end{equation}
where $\epsilon_{i,j,k}=E_{i,j,k}/(E_{i}+E{j}+E_{k})$ are the kinetic energies of each particle, in the reference frame where the emitting source is at rest, normalized to the total energy of the decay. $E_{i,j,k}$ are selected so that $E_{i}\geq E_{j}\geq E_{k}$ and, consequently, $\varepsilon_{i}\geq \varepsilon_{j}\geq \varepsilon_{k}$. In Fig.~\ref{fig:dalitz_plot} we show the Dalitz plot obtained from the experimental data selected with the above discussed procedure (a) compared with the analogous plot constructed with simulated $100\%$ sequential decay (SD) data (b) and the $100\%$ DD$\Phi$ data (c). Simulated data have been obtained with the same prescription used to construct Fig.~\ref{fig:Ex}. In this Dalitz plot representation, a sequential decay (SD) mechanism would populate a uniform horizontal narrow band, while a spread of events along the whole plot region would be observed in the case of DD$\Phi$. The plots of Fig.~\ref{fig:dalitz_plot}(b) and (c) are particularly useful to characterize the expected distortion introduced by the experimental apparatus on the analysis to discriminate the decay mechanism. In particular, two significant conclusions can be extracted from these plots. First, the effect of the detection device on the three $\alpha$ reconstruction results only in a broadening of the SD band, without introducing a significant background contamination in the region outside the band. This result demonstrates that we are able to distinguish between the two mechanisms with an exceptionally low background level. In previous investigations \cite{itoh}, the Dalitz plot constructed with simulated sequential decays shows the presence of data points outside the above mentioned horizontal band, thus containing ambiguities and leading to a reduced sensitivity on direct decay contributions. This difficulties arise from the misassignment of particle tracks inside the strip detectors used in their experiment, as the authors of Ref.~\cite{itoh} state. Our experiment is free from such problems thanks to the use of an hodoscope of independent detectors free of pixel assignment ambiguities. A second, very important, conclusion can be deduced by comparing the behaviour of the experimental Dalitz plot of Fig.~\ref{fig:dalitz_plot}(a) with the simulated ones. An excellent agreement with the simulated SD horizontal band is clearly seen, while only few counts populate the region outside the SD band.

\begin{center}                                                                                                                
	\begin{figure}[t]
		\includegraphics[scale=0.32]{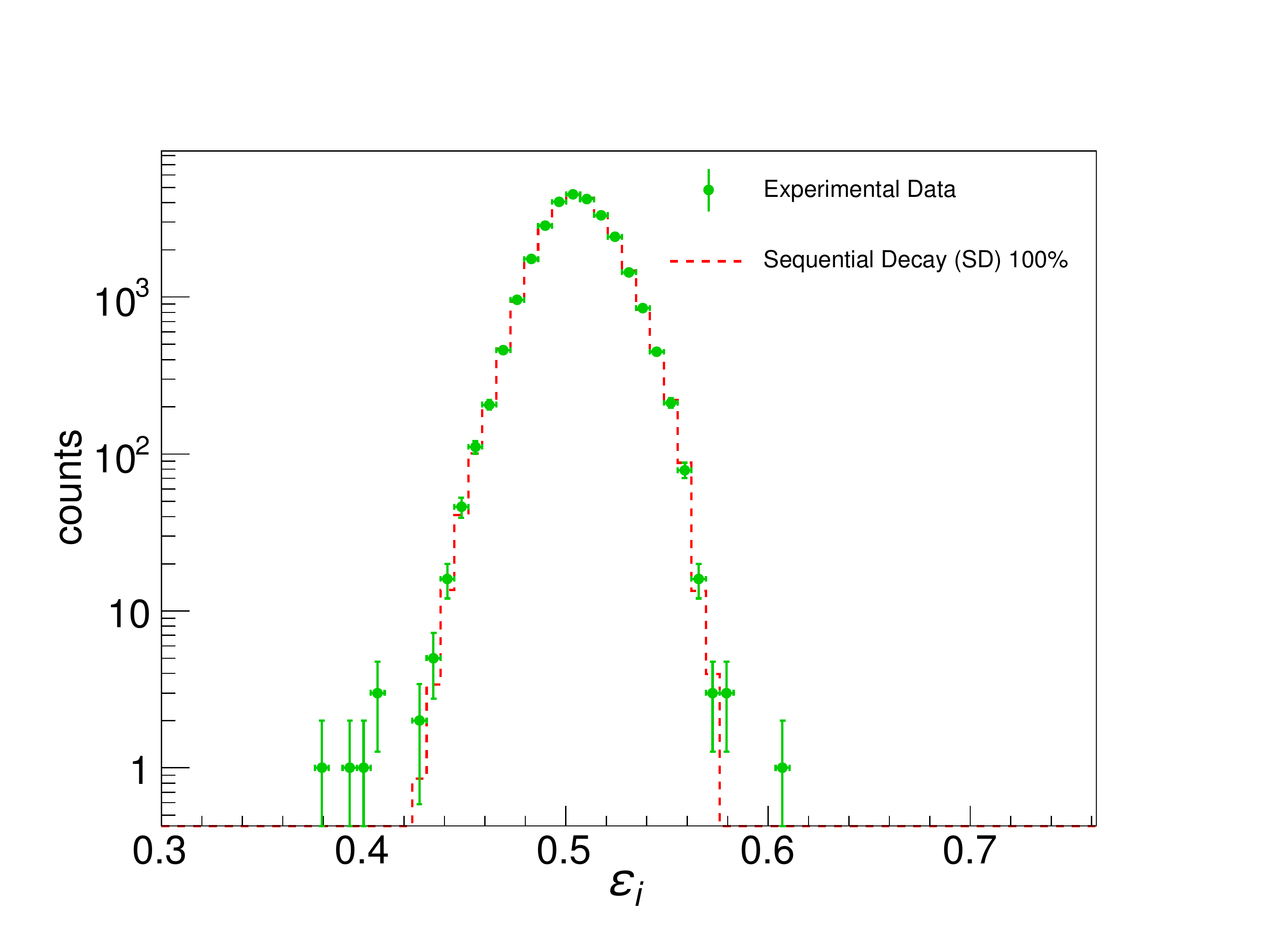}
		\caption{$\epsilon_i$ distribution, i.e. the largest energy among the normalized decay energy of the three $\alpha$ particles in their emitting reference frame. Experimental data (green circles) are compared with the result of a Monte Carlo simulation (red dashed line) where we assumed a $100\%$ sequential decay (SD).}
		\label{fig:epsilon_i}
	\end{figure}	
\end{center}  

A more quantitative analysis can be achieved by inspecting the $\epsilon_i$ distribution, i.e. the distribution of the largest energy among the $\epsilon_{i,j,k}$ normalized energies \cite{itoh}. The $\epsilon_i$ distribution is shown by the green points of Fig.~\ref{fig:epsilon_i}. These values are expected to lie, in the case of a DD$\Phi$ decay, between $0.33$ (when particles share an equal amount of the energy decay) and $0.67$ (when one $\alpha$ is emitted in the opposite direction of the other two). In contrast, a value of about $0.506$ is expected for a SD mechanism. In order to estimate the B.R. of direct decays contributing to the width of the Hoyle state, we have compared the experimental data with the result of a Monte Carlo simulation assuming $100\%$ of SD (red dashed line on Fig.~\ref{fig:epsilon_i}). From an analysis of this spectrum, it is possible to identify an extremely small amount of counts not reproduced by the SD simulation. They correspond to background events falling into the selection of Fig.~\ref{fig:Ex} (the total estimated background level is about $0.036\%$, as previously discussed) and, eventually, to a \emph{signal} of DD.  

Starting from the observed experimental data, we can determine the \emph{lower} and \emph{upper} limits of the DD B.R., by assuming that both the DD and background counts are regulated by the Poisson statistics \cite{statistics}. In doing this evaluation, we follow the Feldman and Cousin's approach to the analysis of small signals described in Ref. \cite{feldman_cousins}, and we carefully take into account the different expected detection efficiencies for DD and SD decay, as determined with Monte Carlo simulations. The lower limit is found to be compatible with \emph{zero}. Therefore we quote an upper limit on the B.R. of the direct three $\alpha$ decay of $0.043\%$ ($95\%$ C.L.). This value is about a factor $5$ lower than the state of art experiment~\cite{itoh}.

To summarize, we have studied the $\alpha$ decay from the Hoyle state ($7.654$, $0^+$) in $^{12}\textrm{C}$ by simultaneously detecting the four $\alpha$ particles emitted from the reaction $^{14}\textrm{N}(\textrm{d},\alpha_2)^{12}\textrm{C}(7.654)$ at an incident energy of $10.5\textrm{MeV}$. To quantitatively estimate the possible contribution of non-resonant (direct) decays bypassing the ground state of $^{8}$Be, we inspect the distribution of the highest normalized energy in the $3\alpha$ decay, $\varepsilon_i$. A complete Monte Carlo simulation, assuming exclusively the sequential decay pattern, fully reproduces the experimental data. The possible presence of any direct decay is found to be statistically insignificant, and an upper limit of $0.043\%$ (C.L. $95\%$) to the corresponding branching ratio is estimated. This finding is in agreement with the previous results by Freer et al.~\cite{freer_hoyle1994}, Kirsebom et al.~\cite{kirsebom} and Itoh et al.~\cite{itoh}, introducing an improvement of about a factor $5$ with respect to the previous most statistically significant work \cite{itoh}. These results provide important information about the $\alpha$ cluster structure of $^{12}$C Hoyle state and have to be carefully taken into account in theoretical models attempting to reproduce the outgoing $\alpha$ particles and the structure of the Hoyle state. They have also very significant astrophysical impact. Indeed, the further reduction of the upper limit of direct decay implies that calculations of the triple-$\alpha$ stellar reaction rate at temperatures lower than $10^8\textrm{K}$ have to be correspondingly revised \cite{nomoto,ogata}.

We gratefully acknowledge all the services (accelerator, target, vacuum lines, mechanics, electronics) of INFN Laboratori Nazionali del Sud (Catania, Italy) for their collective efforts to perform, in the best possible way, the present experiment. We thank the Servizio Elettronica e Rivelatori of the INFN-Sezione di Napoli for the support in the development and production of the hodoscope detector.


\begin{thebibliography}{10}
	
	\bibitem{opik}
	E.J. Opik, Proc. R. Irish Acad. A \textbf{54},  49  (1951).
	
	\bibitem{salpeter}
	E.E. Salpeter, Phys. Rev. \textbf{88},  547  (1952).
	
	\bibitem{hoyle_first}
	F.~Hoyle et~al., Phys. Rev. \textbf{92},  1095c  (1953).
	
	\bibitem{hoyle}
	F.~Hoyle, Astrophys. J. Suppl. Ser. \textbf{1},  121  (1954).
	
	\bibitem{cook}
	C.W. Cook, W.A. Fowler, C.C. Lauritsen, and T.~Lauritzen, Phys. Rev.
	\textbf{107},  508  (1957).
	
	\bibitem{freer_hoyle_review}
	M.~Freer and H.O.U.~Fynbo, Prog. Part. Nucl. Phys.
	\textbf{78},  1  (2014).	
	
	\bibitem{ogata}
	Kazuyuki Ogata, Masataka Kan, and Masayasu Kamimura, Prog. Theor. Phys.
	\textbf{122},  1055  (2009).
	
	\bibitem{Herwig}
	Falk Herwig, Sam~M. Austin, and John~C. Lattanzio, Phys. Rev. C \textbf{73},
	025802  (2006).
	
	\bibitem{Tur:2009zb}
	Clarisse Tur, Alexander Heger, and Sam~M. Austin, Astrophys. J. \textbf{718},
	357  (2010).
	
	\bibitem{nguyen}
	N.B.~Nguyen, F.M.~Nunes and I.J.~Thompson, Phys. Rev. C \textbf{87},
	054615  (2013).
	
	\bibitem{nomoto}
	K.~Nomoto, F.-K. Thielemann, and S.~Miyaji, Astron. Astrophys. \textbf{149},
	239  (1985).
	
	\bibitem{langanke}
	K.~Langanke, M.~Wiescher, and F.K. Thielemann, Z. Physik A - Atomic Nuclei
	\textbf{324},  147  (1986).
	
	\bibitem{nacre}
	C.~Angulo et~al., Nucl. Phys. A \textbf{656},  3  (1999).
	
	\bibitem{nguyenprl}
	N.B.~Nguyen, F.M.~Nunes, I.J.~Thompson and E.F.~Brown, Phys. Rev. Lett. \textbf{109},
	141101  (2012).
	
	\bibitem{garrido}
    E.~Garrido, R.de~Diego, D.V.~Fedorov and A.S.~Jensen, Eur. Phys. J. A \textbf{47}, 102 (2011).	
    
	\bibitem{yabana}
	K.~Yabana and Y.~Funaki, Phys. Rev. C \textbf{85}, 055803  (2012).    
	
	\bibitem{vonoertzen}
	W.~von Oertzen, Zeit. Phys. \textbf{A 357},  355  (1997).
	
	\bibitem{uegaki}
	E.~Uegaki, S.~Okabe, Y.~Abe, and H.~Tanaka, Prog. Theor. Phys. \textbf{57},
	1262  (1977).
	
	\bibitem{kamimura}
	M.~Kamimura, Nucl. Phys. A \textbf{351},  456  (1981).
	
	\bibitem{funaki}
	Y.~Funaki, A.~Tohsaki, H.~Horiuchi, P.~Schuck and G.~R{\"o}pke, Eur. Phys. J. A \textbf{28}, 259 (2006).
	
	\bibitem{funaki1}
	Y.~Funaki, H.~Horiuchi, W.~von~Oertzen, G.~Ropke, P.~Schuck, A.~Tohsaki and T.~Yamada, Phys. Rev. C \textbf{80}, 064326 (2009).
	
	\bibitem{tohsaki}
	A.~Tohsaki, H.~Horiuchi, P.~Schuck, and G.~R\"opke, Phys. Rev. Lett.
	\textbf{87},  192501  (2001).
	
	\bibitem{chernykh}
	M.~Chernykh, H.~Feldmeier, T.~Neff, P.~von Neumann-Cosel, and A.~Richter, Phys.
	Rev. Lett. \textbf{98},  032501  (2007).
	
	\bibitem{morinaga}
	H.~Morinaga, Phys. Rev. \textbf{101},  254  (1956).
	
	\bibitem{epelbaum1}
	E.~Epelbaum et~al., Phys. Rev. Lett. \textbf{106},  192501  (2011).
	
	\bibitem{ishikawa}
	S.~Ishikawa, Phys. Rev. \textbf{C90},  061604  (2014).
	
	\bibitem{freer_hoyle1994}
	M.~Freer et~al., Phys. Rev. C \textbf{49},  R1751  (1994).
	
	\bibitem{raduta}
	Ad.R. Raduta et~al., Phys. Lett. B \textbf{705},  65  (2011).
	
	\bibitem{kirsebom}
	O.S. Kirsebom et~al., Phys. Rev. Lett. \textbf{108},  202501  (2012).
	
	\bibitem{rana}
	T.K. Rana et~al., Phys. Rev. C \textbf{88},  021601(R)  (2013).
	
	\bibitem{morelli_hoyle}
	L.~Morelli et~al., J. Phys. \textbf{G43},  045110  (2016).
	
	\bibitem{itoh}
	M.~Itoh et~al., Phys. Rev. Lett. \textbf{113},  102501  (2014).
	
	\bibitem{freer_models}
    M.~Freer, H.~Horiuchi, Y.~Kanada-En’yo, D.~Lee and Ulf-G.~Mei{\ss}ner, arXiv:1705.06192v1.
	
	\bibitem{koenig}
	W.~Koenig et~al., Il Nuov. Cim. \textbf{39},  9  (1977).
	
	\bibitem{curry}
	J.R. Curry, W.R. Coker, and P.J. Riley, Phys. Rev. \textbf{185},  1416  (1969).
	
	\bibitem{dalitz}
	R.H. Dalitz, Philos. Mag. \textbf{44},  1068  (1953).
	
	\bibitem{statistics}
	R.J. Barlow, {\em Statistics} (J. Wiley \& Sons, Chichester (UK), 1989).
		
	\bibitem{feldman_cousins}
    G.J.~Feldman and R.D.~Cousins, Phys. Rev. D \textbf{57},
	3873  (1998).	

	
\end{thebibliography}
\end{document}